\documentclass[prd,onecolumn,showpacs,groupedaddress,superscriptaddress,amsmath,amssymb,reprint]{revtex4-2}

\usepackage{hhline}
\usepackage{slashed}
\usepackage{graphicx}
\usepackage{graphics}
\usepackage{dcolumn}
\usepackage{bm}
\usepackage{amssymb}
\usepackage{enumerate}
\usepackage{multirow} 

\newcommand*{\Apr}{{A^\prime}}

\def\address{\@ifstar{\address@star}%
  {\@ifnextchar[{\address@optarg}{\address@noptarg}}}

\begin{document}


\title{Improved exclusion limit for light dark matter from $e^+e^-$ annihilation in NA64}
\author{Yu.~M.~Andreev}\affiliation{Institute for Nuclear Research, 117312 Moscow, Russia}
\author{D.~Banerjee}\affiliation{ CERN, European Organization for Nuclear Research, CH-1211 Geneva, Switzerland}
\author{J.~Bernhard}\affiliation{ CERN, European Organization for Nuclear Research, CH-1211 Geneva, Switzerland}
\author{M.~Bond\`i}\affiliation{ INFN, Sezione di Genova, 16147 Genova, Italia}
\author{V.~E.~Burtsev}\affiliation{ Joint Institute for Nuclear Research, 141980 Dubna, Russia}
\author{A.~Celentano}\thanks{Corresponding author}\email{andrea.celentano@ge.infn.it}\affiliation{ INFN, Sezione di Genova, 16147 Genova, Italia}
\author{N.~Charitonidis}\affiliation{ CERN, European Organization for Nuclear Research, CH-1211 Geneva, Switzerland}
\author{A.~G.~Chumakov}\affiliation{ Tomsk Polytechnic University, 634050 Tomsk, Russia}\affiliation{ Tomsk State Pedagogical University, 634061 Tomsk, Russia}
\author{D.~Cooke}\affiliation{ UCL Departement of Physics and Astronomy, University College London, Gower St. London WC1E 6BT, United Kingdom}
\author{P.~Crivelli}\affiliation{ ETH Z\"urich, Institute for Particle Physics and Astrophysics, CH-8093 Z\"urich, Switzerland}
\author{E.~Depero}\affiliation{ ETH Z\"urich, Institute for Particle Physics and Astrophysics, CH-8093 Z\"urich, Switzerland}
\author{A.~V.~Dermenev}\affiliation{ Institute for Nuclear Research, 117312 Moscow, Russia}
\author{S.~V.~Donskov}\affiliation{ State Scientific Center of the Russian Federation Institute for High Energy Physics of National Research Center 'Kurchatov Institute' (IHEP), 142281 Protvino, Russia}
\author{R.~R.~Dusaev}\affiliation{ Tomsk Polytechnic University, 634050 Tomsk, Russia}
\author{T.~Enik}\affiliation{  Joint Institute for Nuclear Research, 141980 Dubna, Russia}
\author{A.~Feshchenko}\affiliation{  Joint Institute for Nuclear Research, 141980 Dubna, Russia}
\author{V.~N.~Frolov}\affiliation{  Joint Institute for Nuclear Research, 141980 Dubna, Russia}
\author{A.~Gardikiotis}\affiliation{ Physics Department, University of Patras, 265 04 Patras, Greece}
\author{S.~G.~Gerassimov }\affiliation{ Technische Universit\"at M\"unchen, Physik  Department, 85748 Garching, Germany}\affiliation{ P.N.Lebedev Physical Institute of the Russian Academy of Sciences, 119 991 Moscow, Russia}
\author{S.~N.~Gninenko}\affiliation{ Institute for Nuclear Research, 117312 Moscow, Russia}
\author{M.~H\"osgen}\affiliation{ Universit\"at Bonn, Helmholtz-Institut f\"ur Strahlen-und Kernphysik, 53115 Bonn, Germany}
\author{M.~Jeckel}\affiliation{ CERN, European Organization for Nuclear Research, CH-1211 Geneva, Switzerland}
\author{V.~A.~Kachanov}\affiliation{ State Scientific Center of the Russian Federation Institute for High Energy Physics of National Research Center 'Kurchatov Institute' (IHEP), 142281 Protvino, Russia}
\author{A.~E.~Karneyeu}\affiliation{ Institute for Nuclear Research, 117312 Moscow, Russia}
\author{G.~Kekelidze}\affiliation{  Joint Institute for Nuclear Research, 141980 Dubna, Russia}
\author{B.~Ketzer}\affiliation{ Universit\"at Bonn, Helmholtz-Institut f\"ur Strahlen-und Kernphysik, 53115 Bonn, Germany}
\author{D.~V.~Kirpichnikov}\affiliation{ Institute for Nuclear Research, 117312 Moscow, Russia}
\author{M.~M.~Kirsanov}\affiliation{ Institute for Nuclear Research, 117312 Moscow, Russia}
\author{V.~N.~Kolosov}\affiliation{ State Scientific Center of the Russian Federation Institute for High Energy Physics of National Research Center 'Kurchatov Institute' (IHEP), 142281 Protvino, Russia}
\author{I.~V.~Konorov}\affiliation{ Technische Universit\"at M\"unchen, Physik  Department, 85748 Garching, Germany}\affiliation{ P.N.Lebedev Physical Institute of the Russian Academy of Sciences, 119 991 Moscow, Russia}
\author{S.~G.~Kovalenko}\affiliation{ Departamento de Ciencias F\'{i}sicas, Universidad Andres Bello, Sazi\'{e} 2212, Piso 7, Santiago, Chile}\affiliation{ Millennium Institute for Subatomic Physics at the High-Energy Frontier (SAPHIR),  ICN2019\_044, ANID, Chile}
\author{V.~A.~Kramarenko}\affiliation{  Joint Institute for Nuclear Research, 141980 Dubna, Russia}\affiliation{ Skobeltsyn Institute of Nuclear Physics, Lomonosov Moscow State University, 119991  Moscow, Russia}
\author{L.~V.~Kravchuk}\affiliation{ Institute for Nuclear Research, 117312 Moscow, Russia}
\author{ N.~V.~Krasnikov}\affiliation{  Joint Institute for Nuclear Research, 141980 Dubna, Russia}\affiliation{ Institute for Nuclear Research, 117312 Moscow, Russia}
\author{S.~V.~Kuleshov}\affiliation{ Departamento de Ciencias F\'{i}sicas, Universidad Andres Bello, Sazi\'{e} 2212, Piso 7, Santiago, Chile}\affiliation{ Millennium Institute for Subatomic Physics at the High-Energy Frontier (SAPHIR),  ICN2019\_044, ANID, Chile}
\author{V.~E.~Lyubovitskij}\affiliation{ Tomsk Polytechnic University, 634050 Tomsk, Russia}\affiliation{ Tomsk State Pedagogical University, 634061 Tomsk, Russia}\affiliation{ Universidad T\'{e}cnica Federico Santa Mar\'{i}a, 2390123 Valpara\'{i}so, Chile}\affiliation{ Millennium Institute for Subatomic Physics at the High-Energy Frontier (SAPHIR),  ICN2019\_044, ANID, Chile}
\author{V.~Lysan}\affiliation{  Joint Institute for Nuclear Research, 141980 Dubna, Russia} 
\author{L.~Marsicano}\affiliation{ INFN, Sezione di Genova, 16147 Genova, Italia}
\author{V.~A.~Matveev}\affiliation{  Joint Institute for Nuclear Research, 141980 Dubna, Russia}
\author{Yu.~V.~Mikhailov}\affiliation{ State Scientific Center of the Russian Federation Institute for High Energy Physics of National Research Center 'Kurchatov Institute' (IHEP), 142281 Protvino, Russia}
\author{L.~Molina Bueno}\affiliation{ ETH Z\"urich, Institute for Particle Physics and Astrophysics, CH-8093 Z\"urich, Switzerland}\affiliation{ Instituto de Fisica Corpuscular (CSIC/UV), Carrer del Catedrátic José Beltrán Martinez, 2, 46980 Paterna, Valencia}
\author{D.~V.~Peshekhonov}\affiliation{  Joint Institute for Nuclear Research, 141980 Dubna, Russia}
\author{V.~A.~Polyakov}\affiliation{ State Scientific Center of the Russian Federation Institute for High Energy Physics of National Research Center 'Kurchatov Institute' (IHEP), 142281 Protvino, Russia}
\author{B.~Radics}\affiliation{ ETH Z\"urich, Institute for Particle Physics and Astrophysics, CH-8093 Z\"urich, Switzerland}
\author{R.~Rojas}\affiliation{ Universidad T\'{e}cnica Federico Santa Mar\'{i}a, 2390123 Valpara\'{i}so, Chile}
\author{A.~Rubbia}\affiliation{ ETH Z\"urich, Institute for Particle Physics and Astrophysics, CH-8093 Z\"urich, Switzerland}
\author{V.~D.~Samoylenko}\affiliation{ State Scientific Center of the Russian Federation Institute for High Energy Physics of National Research Center 'Kurchatov Institute' (IHEP), 142281 Protvino, Russia}
\author{H.~Sieber}\affiliation{ ETH Z\"urich, Institute for Particle Physics and Astrophysics, CH-8093 Z\"urich, Switzerland}
\author{D.~Shchukin}\affiliation{ P.N.Lebedev Physical Institute of the Russian Academy of Sciences, 119 991 Moscow, Russia}
\author{V.~O.~Tikhomirov}\affiliation{ P.N.Lebedev Physical Institute of the Russian Academy of Sciences, 119 991 Moscow, Russia}
\author{I.~Tlisova}\affiliation{ Institute for Nuclear Research, 117312 Moscow, Russia} 
\author{A.~N.~Toropin}\affiliation{ Institute for Nuclear Research, 117312 Moscow, Russia}
\author{A.~Yu.~Trifonov}\affiliation{ Tomsk Polytechnic University, 634050 Tomsk, Russia}\affiliation{ Tomsk State Pedagogical University, 634061 Tomsk, Russia}
\author{P.~Ulloa}\affiliation{ Departamento de Ciencias F\'{i}sicas, Universidad Andres Bello, Sazi\'{e} 2212, Piso 7, Santiago, Chile}
\author{B.~I.~Vasilishin}\affiliation{ Tomsk Polytechnic  University, 634050 Tomsk, Russia}
\author{G.~Vasquez Arenas}\affiliation{ Universidad T\'{e}cnica Federico Santa Mar\'{i}a, 2390123 Valpara\'{i}so, Chile}
\author{P.~V.~Volkov}\affiliation{  Joint Institute for Nuclear Research, 141980 Dubna, Russia}\affiliation{ Skobeltsyn Institute of Nuclear Physics, Lomonosov Moscow State University, 119991  Moscow, Russia}
\author{V.~Yu.~Volkov}\affiliation{ Skobeltsyn Institute of Nuclear Physics, Lomonosov Moscow State University, 119991  Moscow, Russia}
%

\date{\today}


\vskip2.0cm
\begin{abstract}
The current most stringent constraints for the existence of sub-GeV dark matter coupling to Standard Model via a massive vector boson $\Apr$ were set by the NA64 experiment for the mass region $m_\Apr\lesssim 250$ MeV, by analyzing data from the interaction of $2.84\cdot10^{11}$ 100-GeV electrons with an active thick target and searching for missing-energy events. In this work, by including $\Apr$ production via secondary positron annihilation with atomic electrons, we extend these limits in the $200$-$300$ MeV region by almost an order of magnitude, touching for the first time the dark matter relic density constrained parameter combinations. Our new results demonstrate the power of the resonant annihilation process in missing energy dark-matter searches, paving the road to future dedicated $e^+$ beam efforts.
 \end{abstract}


\maketitle
\newpage

The existence of Dark Matter (DM) is proved by multiple, independent astrophysical measurements sensitive to its gravitational effects on ordinary matter. These observations all point to the conclusion that approximately 85$\%$ of the matter of our Universe is made of DM~\cite{Bertone:2016nfn}. Traditionally, most of the experimental DM searches are based on the direct detection of heavy DM particles from the galactic halo, according to the so-called ``WIMP'' scenario~\cite{Arcadi:2017kky}. The current experimental WIMP landscape is controversial~\cite{Cooley:2014aya,Roszkowski:2017nbc}. Despite a few positive observations reported by different collaborations~\cite{Bernabei:2018yyw,Aalseth:2010vx,Agnese:2013rvf,Angloher:2011uu}, the interpretation of these results as a DM signal is in contrast with null measurements reported by other experiments~\cite{Aprile:2018dbl,Akerib:2016vxi,Cui:2017nnn,Agnes:2018fwg,PandaX:2021osp}. Next generation efforts will confirm or rule out this hypothesis~\cite{Aprile:2015uzo,Aalseth:2017fik,Akerib:2015cja,Baudis:2012bc}. 

Motivated by these arguments, in recent years a new alternative hypothesis for the DM nature has been introduced. This predicts the existence of sub-GeV light dark matter (LDM) particles, interacting with SM states through a new force in Nature.  Among the different possibilities, the so-called ``vanilla'' model involves a vector mediator, usually called ``dark photon'' or ``hidden photon'' and denoted as $\Apr$, kinetically mixed with the SM photon. LDM particles are produced via real or virtual $\Apr$ decay~\cite{Holdom:1985ag}. 
The effective Lagrangian density for this model, omitting the LDM mass term, is:
\begin{eqnarray}
{\cal L}_\Apr & \supset & -\frac{1}{4}F^\prime_{\mu\nu} F^{\prime\,\mu\nu} + \frac{\epsilon}{2} F^\prime_{\mu\nu} F^{\mu \nu} + \frac{m^2_{A^\prime}}{2} A^{\prime}_\mu A^{\prime\, \mu} +\nonumber \\
& - & g_D A^\prime_\mu J^\mu_D
\end{eqnarray}
where $F_{\mu\nu}$ and $F^\prime_{\mu\nu}$ are the SM and the dark photon stress tensors, respectively, $J^\mu_{D}$ is the DM  current, $g_D \equiv \sqrt{4\pi \alpha_D}$ is the dark gauge coupling, and $m_{\Apr}$ is the dark photon mass. Finally, $\varepsilon$ is the kinetic mixing parameter between the dark photon and the SM photon, giving rise to an effective $A^\prime$ coupling to SM charged particles. Although the value of the $\varepsilon$ is not predicted by the theory, by making the natural assumption that $g_D \simeq 1$ it is expected that it sits in the interval $\sim 10^{-4} - 10^{-2}$ ($\sim 10^{-6} - 10^{-3}$), if the kinetic mixing is associated to one (two)-loop interactions between the SM and the dark sector~\cite{Essig:2010ye,delAguila:1988jz,ArkaniHamed:2008qp}. We incidentally observe that, equivalently, any new SM extension with an extra $U(1)$ generator that includes a contribution to the hypercharge would result in dark photon coupling of this type~\cite{Fayet:1990wx}.
In this work, we explicitly consider the case $m_\chi < m_\Apr / 2$, where $m_\chi$ is the dark matter particles mass, resulting in an \textit{invisible} decay of the $\Apr$ to LDM particles. This scenario offers a predictive target through a combination of the LDM parameters that is capable of reproducing the presently observed DM relic density~\cite{Boehm:2003hm,Fayet:2004bw}. This can be effectively parameterized in terms of the dimensionless variable $y$ as follows:
\begin{equation}\label{eq:bound}
    y \equiv \alpha_D \varepsilon^2 \left(\frac{m_\chi}{m_\Apr}\right)^4 \rightarrow y \simeq f \cdot 2\cdot 10^{-14}\left(\frac{m_\chi}{1\,\mathrm{MeV}} \right)^2\; \; ,
\end{equation}
where the factor $f$ is a dimensionless $O(1)$ quantity that depends on the fine details of the model.
  
  Experimental searches with accelerators at moderate beam energies (10 GeV -- 100 GeV) have a unique discovery potential in a broad range of the LDM parameter space. Currently, the most stringent exclusion limits in case of an invisibly-decaying dark photon have been reported
  by the NA64 experiment~\cite{NA64:2019imj} for $1$ MeV $\lesssim m_\Apr \lesssim 250$ MeV and by the BaBar experiment~\cite{Lees:2017lec}, for $250$ MeV $\lesssim m_\Apr \lesssim 10$ GeV. A complete review of the current efforts and future proposals, the phenomenological studies associated to this field, and the re-interpretation of past experimental data in this context can be found in Refs.~\cite{Essig:2013lka,Alexander:2016aln,Battaglieri:2017aum,Beacham:2019nyx,Fabbrichesi:2020wbt,Filippi:2020kii,Gninenko:2020hbd,Graham:2021ggy}.

The NA64 experiment at CERN conducts a missing-energy search that exploits a $E_0=100$ GeV high-purity, low-current electron beam from the H4 beamline at CERN North Area impinging on an active thick target. A full description of the NA64 detector and experimental technique can be found, for example, in Refs.~\cite{Gninenko:2016kpg,NA64:2019imj,Gninenko:2013rka,Banerjee:2017hhz}. In the experiment, the momentum of each impinging particle was measured via a magnetic spectrometer consisting of two successive dipole magnets (total magnetic strength $\int B dl \simeq 7$ T$\cdot$ m) and a set of upstream and downstream tracking detectors, Micromegas (MM), Strawtubes (ST) and Gaseous Electron Multipliers (GEM). The overall momentum resolution achieved was $\delta p/p \simeq 1\%$. In order to reduce the intrinsic $1\%$ beam hadron contamination to a negligible level, an ad-hoc beam tagging system based on syncrotron radiation (SR) was developed~\cite{Depero:2017mrr}, using a Pb/Sc sandwich calorimeter to detect the SR photons emitted by the electrons due to their bending in the dipole magnetic field. The active thick target was a 40 radiation length Pb/Sc electromagnetic calorimeter (ECAL), with energy resolution $\sigma_{E}/E\simeq 10\%/\sqrt{E\mathrm{(GeV)}}+4\%$. This was followed by a massive hadronic calorimeter (HCAL), used to detect secondary hadrons and muons produced by the interaction of the primary beam with the target or with other upstream beamline elements. A plastic scintillator-based counter (VETO) was installed between the ECAL and the HCAL to further suppress the background due to muons and other charged particles produced in the ECAL and escaping from it. The trigger for the experiment required the coincidence between the signals of a set of upstream beam-defining plastic-scintillator counters (SC), as well as an in-time cluster in the ECAL with $E_{ECAL} \lesssim  80$ GeV.

The most updated NA64 result corresponds to  $N_{EOT}=2.84\cdot10^{11}$ electrons-on-target (EOT) accumulated during the years 2016, 2017 and 2018. After applying all the selection cuts, determined through a blind-analysis approach by maximizing the experimental sensitivity, no events were found in the signal region, defined by the two requirements $E_{ECAL}<50$ GeV and $E_{HCAL}<1$ GeV. This result was translated to an exclusion limit in the $\Apr$ parameter space - $m_{\Apr}$ vs $\varepsilon$ -, considering only the so-called ``$\Apr$-strahlung'' production mechanism associated with the reaction $e^-Z \rightarrow e^- Z \Apr$, where $Z$ is a nucleus of the active target, followed by the \textit{invisible} $A^\prime$ decay.

In this work, we present a re-evaluation of the LDM exclusion limit from NA64, taking into account for the first time also the $\Apr$ production through the resonant annihilation of secondary positrons of the electromagnetic shower with atomic electrons, $e^+e^-\rightarrow \Apr \rightarrow \chi \overline{\chi}$~\footnote{In this work, we consider the two cases of fermionic and scalar LDM. To not weigh down the notation, in all equations where the final state LDM particles are explicitly reported, we identify these as Dirac fermions whenever this does not introduce any ambiguity.}. 
As discussed in Ref.~\cite{Marsicano:2018glj}, thanks to the resonant cross-section enhancement and to its linear dependence on $\alpha_{EM}Z$, compared to the $\alpha^3_{EM} Z^2$ scaling of the $\Apr$-strahlung reaction (here Z is the charge of a target nuclei), the $e^+e^-$ annihilation channel provides a strong increase to the signal yield, and thus to the exclusion limit, also in case of an electron beam, due to the sizable track length of the secondary positrons in the thick target.

The resonant cross section for a vector $\Apr$ decaying to fermionic or scalar LDM reads:
\begin{equation}
    \sigma_{res}= \frac{4\pi\alpha_{EM}\alpha_D \varepsilon^2}{\sqrt{s}}
    \frac{q\mathcal{K}}{(s-m^2_\Apr)^2+\Gamma_\Apr^2m^2_\Apr \eta}
     \; \; ,
\end{equation}
where $s$ is the $e^+$ $e^-$ system invariant mass squared, $q$ is the LDM daughter particles momentum in the CM frame, and $\Gamma_\Apr$ is the $\Apr$ width, given by 
\begin{eqnarray}
    \Gamma_\Apr&=&\alpha_D\frac{m_\Apr}{3}(1+2r^2)\sqrt{1-4r^2} \,(\mathrm{fermionic\; LDM})\nonumber\\
    \Gamma_\Apr&=&\alpha_D\frac{m_\Apr}{12}(1-4r^2)^{3/2} \; \; (\mathrm{scalar\; LDM}),
\end{eqnarray}
where $r\equiv m_\chi/m_\Apr$, and we neglected the $\varepsilon^2-$suppressed $\Apr$ visible decay channel. Finally, $\mathcal{K}$ is a kinematic factor equal to ${s-4/3q^2}$ (${2/3q^2}$) for the fermionic (scalar) case, while $\eta=(s/m^2_\Apr)^2$ is a correction term introduced for the fermionic LDM case ($\alpha_D=0.5$) to consider the energy dependence of $\Gamma_\Apr$ when this is non-negligible with respect to $m_\Apr$.  

This cross section exhibits a maximum at $s=m^2_\Apr$, i.e. at positron energy $E_R\simeq m^2_\Apr/(2m_e)$. 
By energy conservation, $E_{e^+}\simeq E_\Apr=E_\chi+E_{\overline{\chi}}$: the distribution of the energy sum of the final state LDM pair and, by extension, of the $s-$channel dark photon also shows a maximum at this energy value.

\begin{figure}
    \centering
    \includegraphics[width=.48\textwidth]{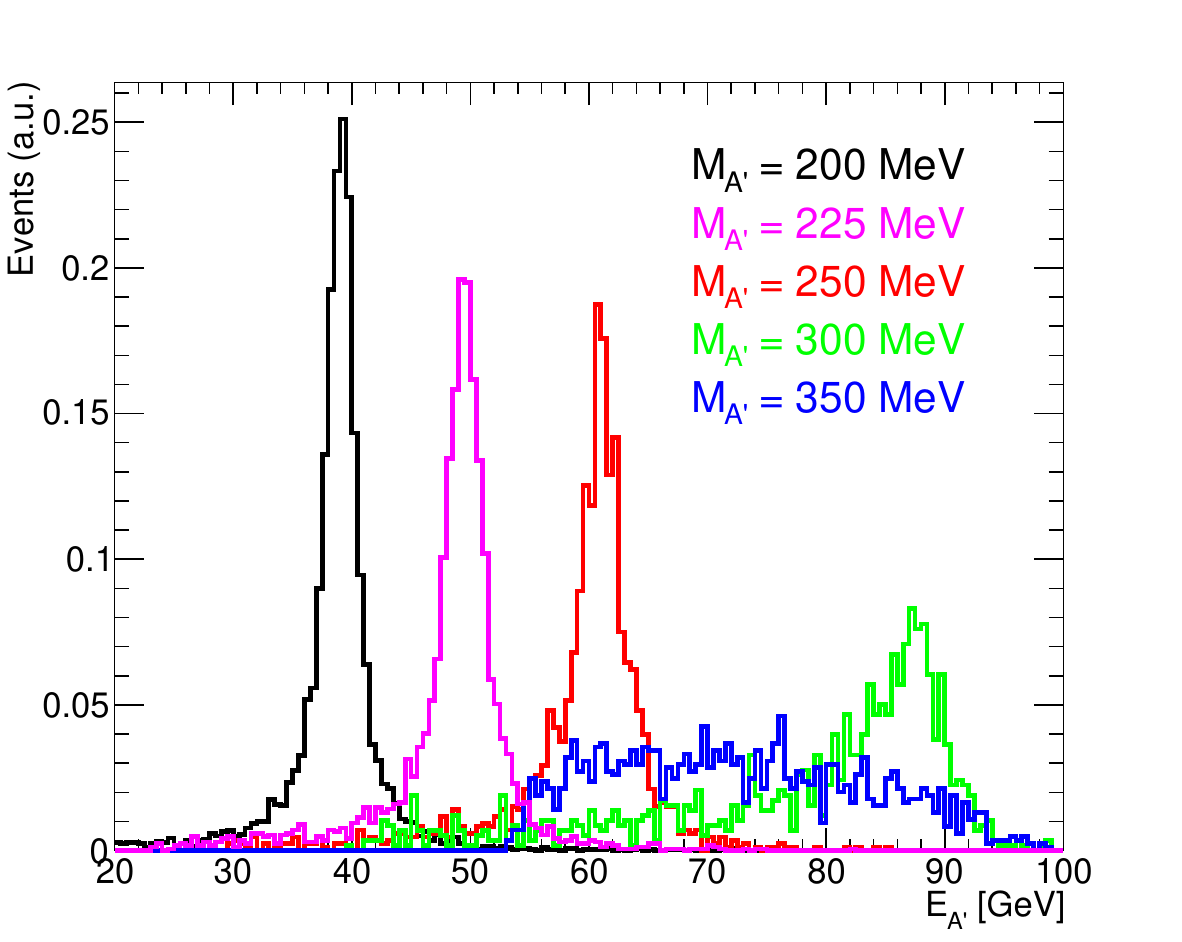}
    \caption{The $\Apr$ energy distribution for $e^+e^-$ annihilation events, for different values of $M_\Apr$. The parameters $\alpha_D=0.1$, $r=1/3$ were used.}
    \label{fig1}
\end{figure}

The expected differential energy distribution of the dark photons produced in the thick target scales as $n(E_\Apr) \propto \sigma_{res}(E_{e^+}) T(E_{e^+})$,  where $T$ is the secondary positrons' track-length distribution~\cite{Chilton,Tsai:1966js,Marsicano:2018glj}. As an example, Fig.~\ref{fig1} shows the $\Apr$ energy distribution for different values of $m_\Apr$ in the fermionic LDM case, fixing $\alpha_D=0.1$ and $r=1/3$. While for low mass values the resonant peak at $E_\Apr=E_R$ is clearly visible, for higher $m_\Apr$ values, corresponding to the case $E_R>E_0$, the dominant contribution to the signal yield, also due to the shape of $T(E_{e^+})$, is associated with the decays of off-shell $\Apr$ produced at the low-energy tail of $\sigma_{res}$ and thus the peak is no longer present. The expected number of signal events with $\Apr$ energy greater than a threshold $E_{miss}^{CUT}$ is given by:
\begin{equation}\label{eq-prod}
N_{Sig}= N_{EOT}\frac{N_A}{A} Z \rho \int_{E_{miss}^{CUT}}^{E_0} dE_{e^+} \,\, T(E_{e^+})\,\widetilde{\sigma}_{res}(E_{e^+}) 
\end{equation}
where $A$, $Z$, $\rho$, are, respectively, the thick target atomic mass, atomic number, and mass density, $N_A$ is Avogadro's number, and $N_{EOT}$ is the number of impinging electrons. Finally, $\widetilde{\sigma}_{res}$ is the annihilation cross section convolved with the active thick target energy resolution. Since the annihilation cross section at the resonance peak reads
\begin{equation}
    \sigma_{res}^P=\frac{1}{\Gamma}\frac{4\pi\alpha_{EM}\epsilon^2}{m_\Apr} \; \; ,
\end{equation}
for a narrow resonance within the interval of energy  accessible by the experiment, i.e. $E^{CUT}_{miss} < E_R < E_0$, the number of expected signal events, roughly scaling as $\sigma_P \cdot \Gamma$, would be, at first order, independent on $\alpha_D$. For larger values of $\alpha_D$, instead, there is a residual dependence due to the actual shape of $\sigma_{res}$. In the analysis we considered separately the two benchmark cases $\alpha_D=0.1$ and $\alpha_D=0.5$, with the fixed mass ratio $r=1/3$~\footnote{It should be noted that the combination $\alpha_D=0.1$ and $m_\chi/m_\Apr=1/3$ has been considered as the benchmark scenario in the recent CERN Physics Beyond Collider report~\cite{Beacham:2019nyx}.}.
Finally, we emphasize that, although the simple fermionic LDM case described previously is already constrained by CMB data~\cite{Madhavacheril:2013cna} for $m_\chi \lesssim$ 10 GeV, it is representative of a set of models involving spin-$\frac{1}{2}$ LDM particles, such as the Majorana or the pseudo-Dirac (small mass splitting) cases~\cite{Akesson:2018vlm}.

This analysis is based on the same dataset already scrutinized for the $\Apr$-strahlung analysis, preventing to adopt a blind analysis approach. Instead, to avoid any bias in the events selection, we decided to conservatively adopt the same cuts employed in Ref.~\cite{NA64:2019imj}. These include the requirements to have, for each event, (I) a single, well identified track in the upstream tracking system, with reconstructed momentum in the window 100 GeV $\pm$ 3 GeV, (II) the energy deposited in the ECAL preshower greater than 0.5 GeV, and (III) the longitudinal and transverse shape of the EM shower compatible with a missing-energy event. We also adopted the same definition for the signal region, identified by the two requirements $E_{ECAL}<50$ GeV, $E_{HCAL}<1$ GeV.
The expected number of backgrounds in the signal region was $(0.53\pm0.17)$, with the largest contribution due to the production of hadrons in the upstream beamline elements by the impinging electron, with the soft $e^-$ measured in the ECAL and the hadrons missed by the HCAL due to insufficient geometric coverage. This estimate is well compatible with the obtained experimental result, corresponding to zero measured events in the signal region.

To evaluate the new exclusion limit as a function of the $\Apr$ mass, we first computed the expected signal yield from the $e^+ e^-$ channel $N_{ann}(\varepsilon_0)$, for the nominal coupling value $\varepsilon_0=10^{-4}$. The signal yield $N_{str}(\varepsilon_0)$ from the ``$\Apr$-strahlung'' mechanism was directly obtained from the published 90$\%$ C.L. NA64 exclusion limit, that corresponds to $N_{up}\simeq2.3$ signal events, via the relation $N_{str}(\varepsilon_0)=N_{up}\cdot \varepsilon_0^2/\varepsilon^2_{UP-str}(m_\Apr)$. The total signal yield was finally computed, $N_{tot}=N_{str}+N_{ann}$, and the new exclusion limit computed as:
\begin{equation}
    \varepsilon^2_{UP}(m_\Apr) = \frac{N_{up}}{N_{tot}} \cdot \varepsilon_0^2
\end{equation}

\begin{table}[b]
    \centering
    \scalebox{.95}{
    \begin{tabular}{c|rrrrr}
         \textbf{Dataset}& \textbf{2016-I} & \textbf{2016-II} & \textbf{2016-III} & \textbf{2017} & \textbf{2018} \\
         \hhline{=|=====}
        EOT ($10^{10}$)&2.3 &1.1 &0.9 & 5.4&18.7 \\
         Efficiency & 0.7 & 0.841 & 0.78 & 0.5 & 0.5 \\
         Eff. uncertainty & $10\%$ & $10\%$ & $15\%$ & $15\%$& $15\%$
    \end{tabular}}
    \caption{The efficiency factors for the different NA64 data sets used in this analysis. See text for details regarding the different procedures used for the 2016 and the 2017-2018 analysis.}
    \label{tab:efficiency}
\end{table}

$N_{ann}(\varepsilon_0)$ was calculated for each $\Apr$ mass by processing a sample of Monte Carlo signal events via the same NA64 reconstruction code used for the data analysis. We employed a GEANT4-based simulation~\cite{Agostinelli:2002hh} of the NA64 setup, using the DMG4 package~\cite{Celentano:2021cna} for events generation. To optimize the simulation time, an ad-hoc cross-section biasing scheme was implemented. We set to zero the production cross section below a certain impinging positron energy $E^{prod}_{cut}=42.5$ GeV, to avoid the production of signal events with a low-energy $\Apr$, that would not satisfy the ECAL missing energy cut, even considering the finite ECAL energy resolution. Then, we artificially enhanced the production cross section above $E^{prod}_{cut}$, multiplying it by a constant factor $\beta$, tuned independently for each mass value to avoid double-counting effects.

We explicitly took into account additional additional efficiency corrections  for effects that are not included in the  simulation package, depending on the different run periods. For the three 2016 datasets the efficiency corrections were determined by comparing the measured di-muon yield with the one predicted by the Monte Carlo simulations (see Ref.~\cite{Banerjee:2017hhz}, Sec. VI). Comparing the yield and the distribution of events between data and Monte Carlo, the efficient corrections factors for the trigger, the SRD, and the ECAL selections, as well for the DAQ dead-time, were determined, together with the corresponding uncertainty. Effects due to the SRD cut and the VETO cut were taken into account by applying them also to the simulated Monte Carlo events. Further studies were performed exploiting data collected with an ``open-trigger'' configuration, without the ECAL energy cut, to determine the VETO and the HCAL selection signal efficiency corrections, that were found to be compatible with one. For the 2017 and 2018 datasets, instead, a slightly different procedure was used. Monte Carlo events were reconstructed using a loose set of cuts, that include the ECAL and the PRS thresholds only. The corrections due to the other cuts employed in the analysis were extracted from the data collected with the ``open-trigger'' configuration, corresponding to an almost pure sample of 100-GeV electrons impinging on the detector. The efficiency associated to each subdetector was determined from the fraction of events satisfying the corresponding cut~\cite{NA64:2019imj}.


\begin{figure}
    \centering
    \includegraphics[width=.46\textwidth]{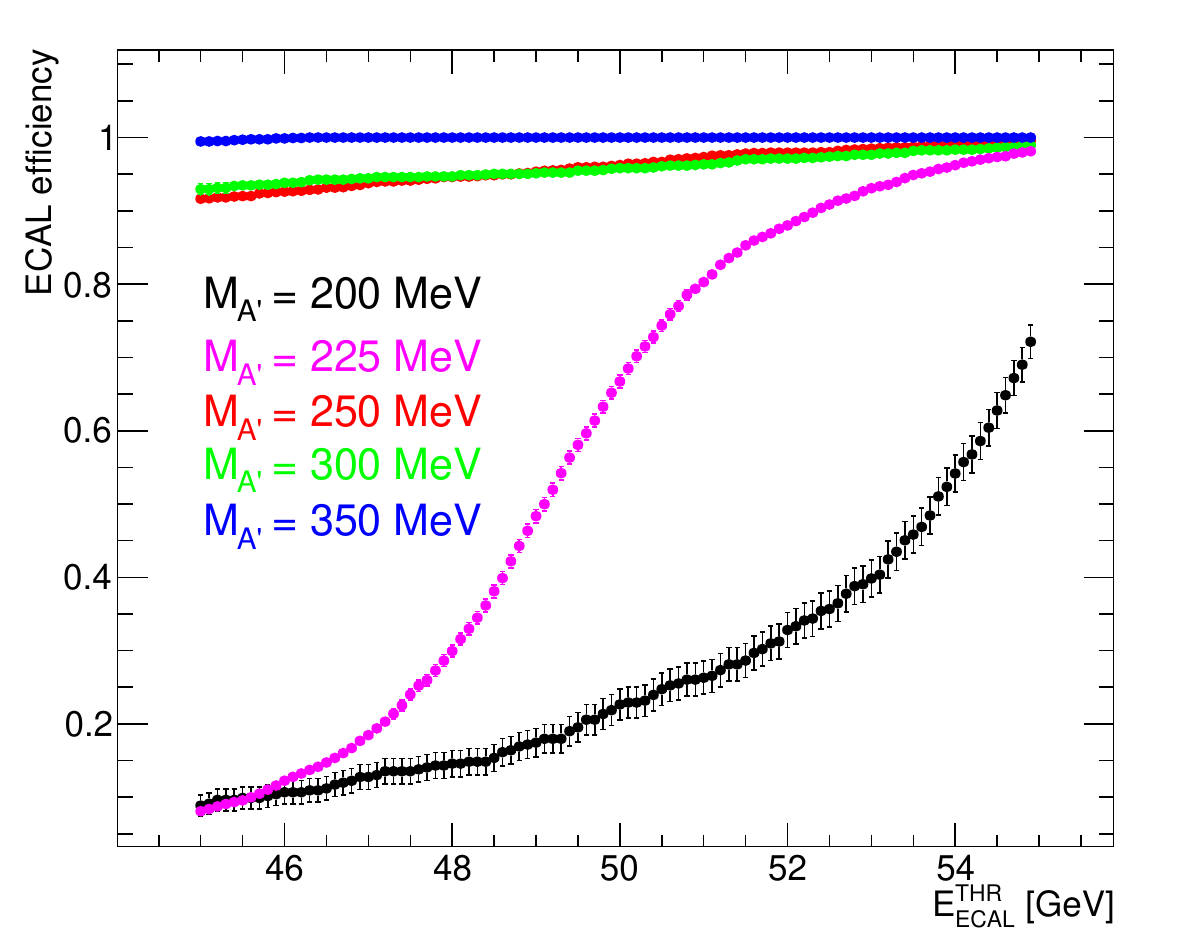}
    \caption{The ECAL signal efficiency curve for LDM production via $e^+e^-$ annihilation as a function of $E_{ECAL}^{THR}$, for different values of the dark photon mass.}
    \label{fig2}
\end{figure}

The results are summarized in Table I. Although the different procedures used to determine them for the 2016 and the 2017/2018 datasets does not allow for a direct comparison, taking into account all effects the overall efficiency for the high-intensity periods, of about $5.5 \cdot 10^6$ e$^-$/spill (2017) and $7 \cdot 10^6$ e$^-$/spill (2018), compared to $3.7 \cdot 10^6$ e$^-$/spill for 2016, is approximately 10$\%$ lower, mostly due to pile-up effects. The efficiencies uncertainties account for effects that are the same between the original $\Apr$-strahlung only analysis and this work. These include the uncertainty associated with the trigger, the tracking, the SRD, the VETO, and the HCAL subsystem, and to the corrections due to pile-up. The dominant uncertainty factor, of the order of $10\%$, was associated to the difference between the predicted and the measured dimuon signal yield. To further account for the significantly different $E_{\Apr}$ distribution associated to the $\Apr-$strahlung and $e^+e^-$annihilation processes, we computed separately, for the latter channel, the systematic uncertainty associated with a possible shift in the ECAL absolute energy scale, by means of Monte Carlo simulations, evaluating the corresponding signal efficiency curve as a function of the ECAL threshold $E_{ECAL}^{THR}$ ($E_{ECAL}^{THR}=E_0-E_{miss}^{CUT}$). To properly sample the $\Apr$ production, for this study we lowered $E^{prod}_{cut}$ to 20 GeV. The obtained result is shown in Fig.~\ref{fig2}. As expected, the steepest curve is seen for $m_\Apr \simeq 225$ MeV, since in this case the resonant energy corresponds to the nominal 50 GeV ECAL missing energy threshold. The uncertainty on the ECAL energy scale is mostly due to short-term  fluctuations within individual SPS spills of the ECAL PMTs gain that are not corrected for in the calibration procedure. This effect was quantified using data collected during the 2018 high-intensity run period with the ``open-trigger'' configuration, tracking the position of the 100 GeV deposition peak as a function of the event time relative to the beginning of the spill, and found to be approximately 3$\%$. This corresponds to a $\pm 1.5$ GeV variation of $E^{THR}_{ECAL}$ that translates to a $\simeq35\%$ uncertainty on the signal efficiency at this mass value, already dropping to $1.5\%$ for $m_\Apr = 250$ MeV.
A similar procedure, applied to the ECAL preshower threshold, showed that the corresponding signal efficiency was approximately $100\%$ with negligible systematic uncertainty.

\begin{figure}
    \includegraphics[width=.52\textwidth]{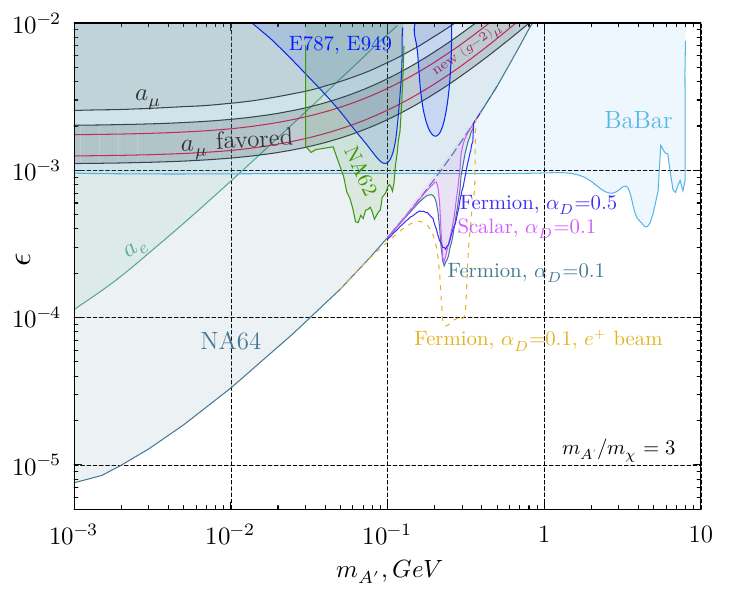}
    \caption{The new NA64 exclusion limit including the resonant $A^\prime$ via $e^+e^-$ annihilation, comparing the scalar and the fermionic LDM cases. Existing limits from BaBar~\cite{Lees:2017lec}, E787 and E949~\cite{Essig:2013vha}, and NA62~\cite{CortinaGil:2019nuo} are shown, as well as the favored area from the muon $g-2$ anomaly~\cite{Fayet:2007ua}, also including the new result that takes into account the latest results from Fermilab~\cite{Abi:2021gix} (red lines). The dashed cyan line report the previous NA64 result, without including the contribution from $e^+e^-$ annihilation. The orange dashed line is the sensitivity projection for a NA64-like experiment with an $e^+$ beam, assuming the same run conditions and accumulated statistics.}
    \label{fig3}
\end{figure}

\begin{figure*}[t]
    \includegraphics[width=.48\textwidth]{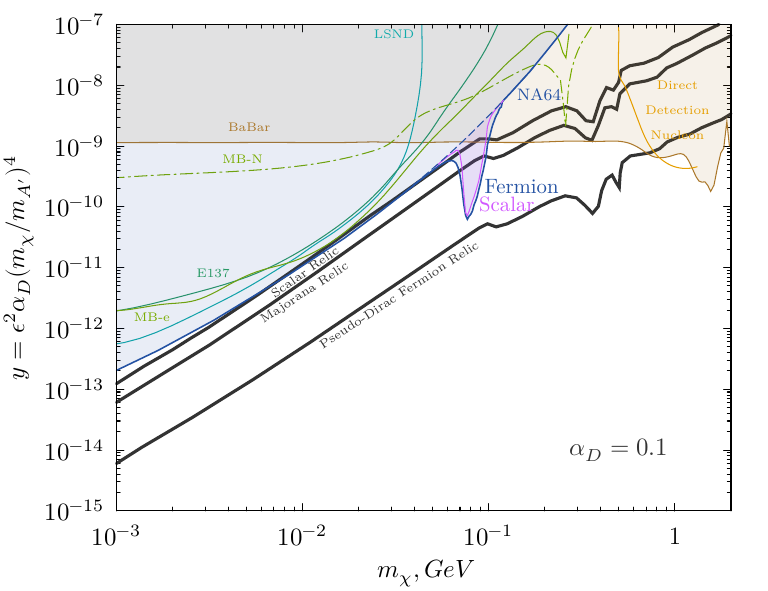} 
    \includegraphics[width=.48\textwidth]{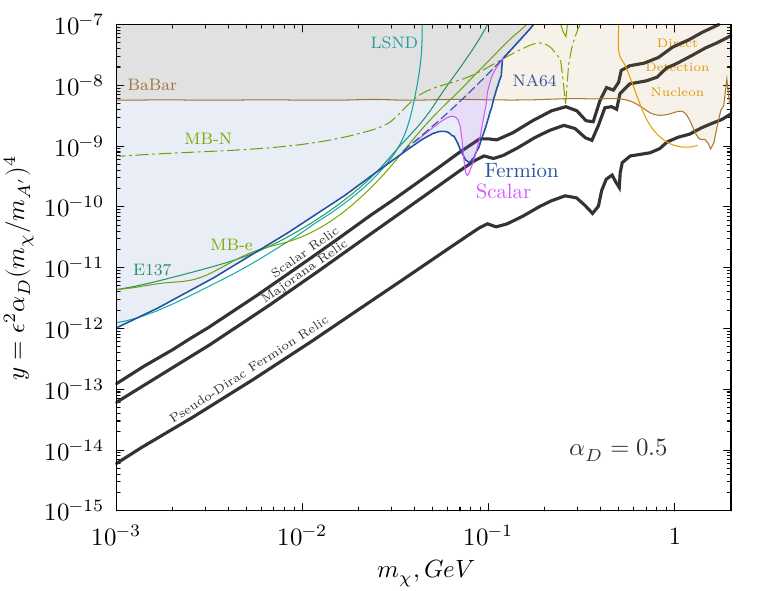}
    \caption{The new NA64 exclusion limit in the $(y,m_\chi)$ plane, including the $e^+e^-$ annihilation process, in the $(m_\chi,y)$ plane, for $\alpha_D=0.1$ (left) and $\alpha_D=0.5$ (right). The other curves and shaded areas report already-existing limits in the same parameters space from E137~\cite{Batell:2014mga}, LSND~\cite{deNiverville:2011it,Batell:2009di}, MiniBoone~\cite{Batell:2009di}, and BaBar~\cite{Lees:2017lec}.
    The black lines show the favored parameter combinations for the observed dark matter relic density for different variations of the model.}
    \label{fig4}
\end{figure*}

The new exclusion limit in the $\Apr$ parameter space ($\varepsilon$ vs $A^\prime$ mass) are shown in Fig.~\ref{fig3}, for the two model variations discussed before. We observe that, due to the significantly smaller $\Apr$ width predicted by the scalar case, in this case the shape of $\widetilde{\sigma}_{RES}$ doesn't change significantly with $\alpha_D$, resulting to almost the same exclusion limit for the two values $\alpha_D=0.5$ and $\alpha_D=0.1$. Thanks to the signal yield enhancement provided by the resonant annihilation mechanism, the new limit is up to one order of magnitude stronger than the currently published $A^\prime$-strahlung limit in the mass range between 200 and 300 MeV, corresponding to a resonant energy approximately between 40 GeV and 90 GeV. The sensitivity projection for a NA64-like experiment performed with a 100 GeV positron beam, assuming the same run conditions and accumulated statistics, is also reported for comparison in the same Figure, to highlight the strong potential of an $e^+$-beam effort in exploring the large-mass $\Apr$ region. The possibility to perform such a measurement in the future with the NA64 experiment is currently under evaluation within the collaboration.

As discussed before, this result was obtained without including explicitly any systematic uncertainty in the upper limit evaluation. To check the effect of this procedure, we performed a full re-evaluation of the experimental upper limit for the worst case scenario $M_{A^\prime}=225$ MeV and $\alpha_D=0.1$, analyzing simultaneously the 2016, 2017 and 2018 datasets using the multibin limit setting technique described in Ref.~\cite{Banerjee:2017hhz}, adding the contributions from the $\Apr-$strahlung and the $e^+e^-$-annihilation in the expected signal yield. The systematic uncertainties associated with the efficiency corrections discussed before, the background estimate, and the EOT number ($\pm 5\%$) were added as independent nuisance parameters in the likelihood model, with a log-normal distribution~\cite{Gross:2007zz}. The dominant factor affecting the upper limit value is the efficiency correction uncertainty, of about $35\%$ ($10\%$) for the $e^+e^-$ ($\Apr$-strahlung) channel. The obtained upper limit for $\varepsilon$ was $10\%$ lower than the one calculated with the simplified procedure discussed previously. Since for different $m_\Apr$ values the ECAL threshold effect on the $\Apr$ signal eefficiency is significantly smaller, as discussed previously, we decided to conservatively quote the results obtained from the latter, and to consider the $10\%$ variation as a worst-case estimate of the systematic uncertainty associated with the limit extraction procedure itself.

These new results were also used to derive exclusion LDM limits  the $y$ vs $m_\chi$ parameter space, reported in Fig.~\ref{fig4} for $\alpha_D=0.1$ (left panel) and $\alpha_D=0.5$ (right panel), together with the already-excluded regions from other experiments, and with the so-called ``thermal-targets'', i.e. the preferred combination of the parameters to explain the observed dark matter relic density, considering different variations of the model. These bounds were calculated through Eq.~\ref{eq:bound} using the same procedure adopted in Ref.~\cite{NA64:2019imj}. Our new results robustly exclude, for the first time, the region of the LDM parameters space extending to the scalar and Majorana fermion ``thermal-target'' lines for the LDM mass range between $70$ MeV and $95$ MeV for $\alpha_D=0.1$. For $\alpha_D=0.5$, instead, only the scalar ``thermal target'' region between $70$ MeV and $90$ MeV is excluded, while the Majorana thermal target is just touched for $m_\chi=80$ MeV.

In conclusion, we extended the existing NA64 exclusion limit for an invisibly-decaying dark photon by considering the production channel associated with the resonant annihilation of secondary positrons with atomic electrons. This mechanism was actually found to be the dominant one for the $A^\prime$ mass range between 200 MeV and 300 MeV, allowing us to set more stringent limits in the LDM parameters space, touching for the first time the ``thermal-target'' lines for scalar and Majorana fermion models between $70$ MeV and $95$ MeV. Looking forward, we expect to further exploit the $e^+e^-$ annihilation production mechanism in the future NA64 data-taking runs by considering it at the earliest stage of the analysis, before data unblinding, together with the $\Apr$-strahlung channel during the signal window definition process.

\begin{acknowledgments}
We gratefully acknowledge the support of the CERN management and staff and the technical staffs of the participating institutions for their vital contributions. 
This result is part of a project that has received funding from the European Research Council (ERC) under the European Union’s Horizon 2020 research and innovation programme, Grant agreement No. 947715 (POKER). This work was supported by the HISKP, University of Bonn (Germany), JINR (Dubna), MON and RAS (Russia), ETH Zurich and SNSF Grant No. 169133, 186181, 186158, 197346 (Switzerland), and grants FONDECYT 1191103, 1190845, and 3170852, UTFSM PI~M~18~13 and ANID PIA/APOYO AFB180002 and ANID - Millenium Science Initiative Program - ICN2019\_044 (Chile).
\end{acknowledgments}

%


\end{document}